\documentclass[useAMS,,letters,usegraphicx]{mn2e}
\usepackage{amssymb}
\usepackage{epsf}
\usepackage{xspace}
\usepackage{times,amsmath}

\newcommand{\etal}{et al.\ }
\newcommand{\beq}{\begin{equation}}
\newcommand{\eeq}{\end{equation}}

\newcommand{\aap}{A\&A}
\newcommand{\apj}{ApJ}

\title[Self-Consistent Analysis of OH Zeeman Observations]
{Testing Molecular-Cloud Fragmentation Theories: Self-Consistent Analysis 
of OH Zeeman Observations}
\author[T. Ch. Mouschovias and K. Tassis]{Telemachos Ch. Mouschovias$^{1}$ and Konstantinos Tassis$^{2}$ \\
$^{1}$Departments of Physics and Astronomy, University of Illinois at 
Urbana-Champaign, 1002 West Green Street, Urbana, IL 61801\\
$^{2}$Jet Propulsion Laboratory, California Institute of 
Technology, Pasadena, CA, 91109}
\begin{document}
\date{}
\maketitle

\begin{abstract}

The ambipolar-diffusion theory of star formation predicts the formation 
of fragments in molecular clouds with mass-to-flux ratios greater than that 
of the parent-cloud envelope. By contrast, scenarios of turbulence-induced 
fragmentation do not yield such a robust prediction. Based on this 
property, Crutcher et al. (2009) proposed an observational test that could 
potentially discriminate between fragmentation theories. However, the analysis 
applied to the data severely restricts the discriminative power of the test: 
the authors conclude that they can only constrain what they refer to as the
``idealized'' ambipolar-diffusion theory that assumes initially straight-parallel 
magnetic field lines in the parent cloud. We present an original, self-consistent 
analysis of the same data taking into account the nonuniformity of the magnetic 
field in the cloud envelopes, which is suggested by the data themselves, and we 
discuss important geometrical effects that must be accounted for in using this 
test. We show quantitatively that the quality of current data does not allow for 
a strong conclusion about any fragmentation theory. Given the discriminative 
potential of the test, we urge for more and better-quality data.
\end{abstract}


\section{Introduction}\label{intro}

The ratio of the mass and magnetic flux of interstellar molecular clouds 
has received well-deserved observational attention in recent years (e.g., 
Crutcher 1999; Heiles \& Crutcher 2005). For a cloud as a whole, the 
mass-to-flux ratio is an important {\em input} to the ambipolar-diffusion 
theory of fragmentation (or core formation) in molecular clouds (e.g., see
Fiedler \& Mouschovias 1992, eq. [8]; 1993, eq. [1c] and associated 
discussion). What the ambipolar-diffusion theory {\it predicts} is the 
mass-to-flux ratio of {\it fragments} (or {\it cores}) in molecular clouds 
and how this quantity evolves in time from typical densities $\simeq 10^{3} \, 
{\rm cm^{-3}}$ to densities $\simeq 10^{14} \, {\rm cm^{-3}}$ (Tassis 
\& Mouschovias 2007). Observations have been in excellent quantitative 
agreement with the theoretical predictions in that the mass-to-flux ratio {\it 
of cores} is found to be supercritical by a factor 1 - 4 (Crutcher \etal 1994; 
Crutcher 1999 and correction by Shu \etal 1999, pp. 196 - 198; Ciolek \& 
Basu 2000; Troland \& Crutcher 2008; Falgarone et al. 2008). By contrast, 
simulations of turbulence-driven fragmentation do 
not find cores with systematically greater mass-to-flux ratios than those
of their parent clouds (e.g., Lunttila \etal 2008). Therefore, the effort by 
Crutcher \etal (2009) (hereinafter CHT) to measure the variation of the 
mass-to-flux ratio from the envelopes to the cores of four molecular clouds 
and thereby constrain cloud-fragmentation theories is a much needed 
observational test.

The effort by CHT to measure the magnetic field in four cloud envelopes 
yielded mostly nondetections, allowing only the placement of weak upper 
limits. Also, the data are suggestive of spatial variations of the field 
in the cloud envelopes. This spatial variation must be explicitly 
treated in the data analysis. Instead, CHT performed an analysis 
based on the overly restrictive (and contradicted by the data) assumption of 
uniform magnetic field in the envelope, which minimizes the potentially 
constraining power of their observations. CHT attempt to justify their 
restrictive assumption by claiming that they are testing the ``idealized 
ambipolar-diffusion model'' that assumes initially straight-parallel field 
lines in the parent cloud. Thus, if the data and the data analysis in CHT 
are taken at face value, they at best test an {\em input} to a theory, not 
the prediction of the theory relating to the variation of the mass-to-flux 
ratio from a core to its envelope, {\em given} the field strength and its 
spatial variation in the envelope. As we show below, the geometry of the 
field lines in a parent cloud crucially affects the 
{\it observed} variation of the mass-to-flux ratio from a core to the 
envelope while the fundamental prediction of the ambipolar-diffusion 
theory (that the mass-to-flux ratio increases from the envelope to the core) 
remains unchanged.

In this letter, we present a novel analysis of the OH-Zeeman data, 
applicable also to other sets of data that show intrinsic variation of 
the quantity being measured. 

\section{Data Analysis}\label{analysis}

The CHT data consist of existing OH Zeeman measurements in four molecular 
cloud cores and of four new measurements in the region surrounding each 
of these cores (in the clouds L1448, B217-2, L1544, and  B1). For each 
observation of an envelope's line-of-sight magnetic field $B_j$, CHT 
quote an associated Gaussian uncertainty $\sigma_j$. These four values 
for each cloud envelope are shown in Table \ref{crutable}, col. 2 - 5 
(taken from CHT Figs. 2 - 5). 
%
\begin{table}
\begin{center}
\caption{\label{crutable} Magnetic Fields and Errors (in $\mu$G) in Four Cloud 
Envelopes (data from CHT).}
\begin{tabular}{c|rrrr}
\hline \hline 
Cloud & $B_1\pm\sigma_1$ & $B_2\pm\sigma_2 $& $B_3\pm\sigma_3 $ & $B_4\pm\sigma_4$ \\
\hline
L1448CO & $-9\pm 13$ & $-11\pm 6$ & $-7 \pm 7$ & $14\pm 8$ \\
B217-2 & $-13\pm 9$ & $5 \pm 6$ & $6 \pm 8$ & $9 \pm 13 $ \\
L1544 & $-3 \pm 4$ & $-1 \pm 4$ & $22 \pm 6$ & $2 \pm 10$ \\
B1 & $-16 \pm 6$ & $0 \pm 7$ & $-3 \pm 6$ & $-10 \pm 5$\\ 
\hline \hline
\end{tabular}
\end{center}
\vspace{+3ex}
\end{table}

\subsection{The CHT Analysis}\label{CHT}

CHT assign a value to the magnetic field strength in each envelope, which is
obtained from a simultaneous least-squares fit over the 8 Stokes V spectra 
(2 spectral lines at each of 4 positions in each envelope). The fit gives a 
{\em single value} of the line-of-sight field and a single value  of its 
uncertainty in each envelope. The uncertainty was calculated under the assumption 
that there is no intrinsic spatial variation of the field strength in each 
cloud envelope and, therefore, any spread in the observed $B_j$ values is 
attributed to observational errors. The CHT values for the envelope fields and 
their uncertainties are shown in Table 2, column 2.

Using this mean field, CHT calculate what they regard as the magnetic flux 
of the envelope, which, combined with the flux in the core, is used to obtain 
the quantity $R$ defined by 
$R=\left(I_{\rm core} \Delta V_{\rm core}/{B_{\rm core}}\right)/
\left(I_{\rm env}\Delta V_{\rm env}/{B_{\rm env}}\right)$. The quantity
$I$ is the peak intensity of the spectral line in degrees K, $\Delta V$ is 
the FWHM in km $\rm s^{-1}$, and $B$ is the line-of-sight mean field in $\mu$G. 
A value $R=1$ would imply that the mass-to-flux ratio does not vary from 
an envelope to a core in the same cloud, while $R\geq 1$ would imply a 
mass-to-flux ratio greater in the core than in the envelope. Since most of the 
CHT measurements of $B$ in cloud envelopes are nondetections, the analysis 
relies sensitively on the treatment and propagation of observational 
uncertainties to obtain limits on the derived quantity $R$.  

As mentioned above, CHT calculate a mean value of $B_{\rm env}$ and an
uncertainty on this mean {\em under the explicit assumption} that the
magnetic field in the envelope can be described by a {\em unique}
$B_{\rm env}$ value, which their analysis seeks to constrain. However, 
the magnetic field in the cloud envelope is not known {\it a priori}  
to have a unique uniform value. In fact, the data suggest the opposite 
(e.g., observations 2 and 4 in L1448CO differ by more than 3$\sigma$; observations 
1 and 3 in B217-2 differ by more than $2\sigma$; observations 1 and 3 in L1544 differ 
by more than 4$\sigma$; observations 1 and 2 in B1 differ by more than $2\sigma$; 
see Table\ref{crutable}). CHT justify this choice by restricting their comparison 
to what they call the ``idealized'' ambipolar-diffusion theory, assuming that the 
field lines in the molecular cloud envelope are straight and parallel. 

If, as the CHT data suggest, the assumption of zero-spread $B_{\rm env}$ 
is relaxed, the uncertainties CHT calculate are not the relevant ones. 
A simple example will illustrate the point: Consider a cloud envelope in 
which the magnetic field has a distribution of values with mean 10 $\mu$G 
and spread 5 $\mu$G. An observer makes only two measurements of the envelope 
field, each with uncertainty 0.1 $\mu$G. The first measurement gives 
$10 \, \pm 0.1 \, \mu$G, and the second measurement gives $14 \, \pm \, 0.1 \, 
\mu$G (both very likely). Under the CHT assumption of zero spread, the mean 
and associated uncertainty are simply the average, $B_{\rm mean} = 12 \, \mu$G, 
and the propagated observational error, 
$\sigma_{\rm mean} = {(\sum_{j=1}^2 \sigma_j^2)^{1/2}}/2 = 0.07 \, \mu$G.
Clearly, however, this $B_{\rm mean}$ differs from its true value by $2 \, \mu$G, 
not by $0.07 \, \mu$G. In other words, if there is significant spatial 
variation of $B$ in a cloud envelope, the CHT-kind of analysis grossly 
underestimates the uncertainty on the mean. 

\subsection{Straight-Parallel Field Lines in Cloud Envelopes?}

A simple inspection of the CHT raw data, taken at face value, reveals that 
these four clouds do not have straight-parallel field lines in their envelopes. 
But are such clouds expected on the basis of theoretical considerations? 
Straight-parallel field lines in a parent cloud is an idealization in some 
theoretical calculations that renders a mathematically complicated multifluid, 
nonideal MHD system tractable while capturing all the essential physics of the 
core formation and evolution problem. However, it has never been suggested that 
in a {\em real} cloud, which is an integral part of a dynamic ISM, the envelope 
field lines will be straight and parallel. Distortions superimposed on the 
characteristic hour-glass morphology associated with the compression of the 
field lines during gravitational core formation are routinely expected. 

\begin{figure}
\label{FLs}
\begin{center}
\includegraphics[width=2.5in]{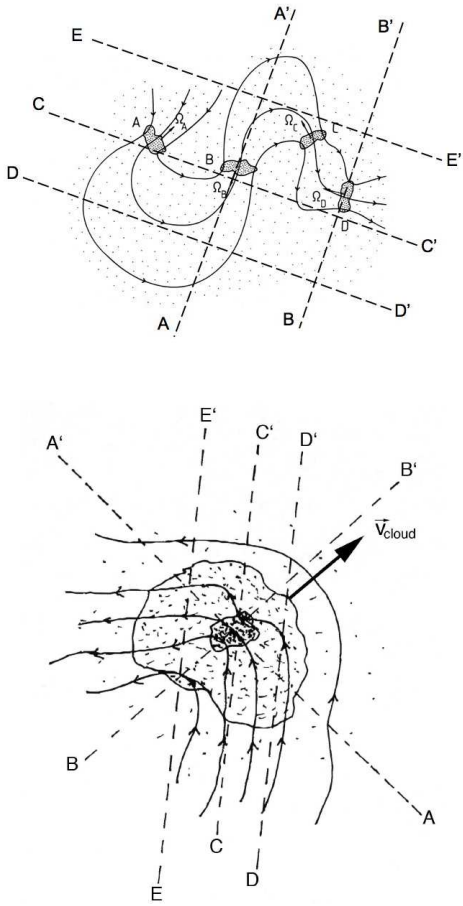}
\end{center}
\caption{Schematic diagrams: ({\it a, top}) A deformed flux tube that has 
fragmented along its length in a molecular cloud (from Mouschovias \& Morton 1985). 
The deformation can be caused by the relative motion of the fragments. 
({\it b, bottom}) Deformation of the field lines threading a cloud caused by its 
motion relative to the surrounding medium. The cloud is shown, for simplicity, to 
contain only one fragment (or core), in the neighborhood of which the hourglass 
shape of the field lines had been established during core formation but affected 
by the cloud's motion.The dashed lines in both figures represent different lines 
of sight, whose significance is explained in the text.}
\end{figure}

Mouschovias \& Morton (1985, Fig. 13) had sketched what they regarded as a 
more realistic field geometry in a molecular cloud in which there are several 
(in that case four) magnetically connected fragments. That figure is reproduced 
here as Figure 1a. This configuration can result from relative motion of the 
fragments (labeled A, B, C, and D) within the cloud, due to the cloud's mean
gravitational field. The motion of a cloud as a whole relative to the intercloud 
medium will also bend the magnetic field lines in an almost U-shape, as shown in 
Figure 1b. One can easily visualize lines of sight in Figures 1a and 1b (e.g., the 
line CC$^{\prime}$) along which a measurement would yield $B_{\rm los,env} > 
B_{\rm los,core}$, although the actual field strength in the core is greater 
than that in the envelope as evidenced by the compressed field lines in the core. 
By contrast, along AA$^{\prime}$, almost the full strength of the core's magnetic 
field will be measured, but only a fraction of the envelope's field strength will 
be detected. Altogether: (1) An idealization in a theoretical calculation should 
not be mistaken for a prediction. (2) Observations that may potentially reveal the 
geometry of the field lines can and should be used as {\em input} to build a 
particular model for the observed cloud (as done in the case of B1 by Crutcher 
\etal 1994, and for L1544 by Ciolek \& Basu 2000). (3) The geometry of the field 
lines cannot be ignored in analyzing data from observations that measure only one 
component of the magnetic field (e.g., Zeeman observations) if the purpose is to 
test a theory or discriminate between alternative theories. The new analysis of 
the CHT data in \S~2.4 accounts for the field geometry suggested by the data 
themselves. 

\subsection{Using Data to Calculate a Cloud's Magnetic Flux}\label{geometry}

Unlike the CHT analysis, if the data show field reversals, the positive and 
negative values of the measured $B_{\rm los, env}$ must not be algebraically 
averaged (which is what the CHT assumption of a single magnetic field value in 
the envelope imposes on the data) and then multiplied by the plane of the sky 
area of the envelope in order to obtain its magnetic flux. If the three (perhaps 
all four) of the observed cloud envelopes exhibited true reversals in the field 
direction (but see \S~2.4 below), that would imply a bent magnetic flux tube 
threading each cloud. In such a case, {\em only one} algebraic sign of the 
magnetic field (the one corresponding to the greatest absolute values) should 
be considered in estimating the magnetic flux of the envelope. Figure 2 and its 
caption clarify this point. 

\begin{figure}
\label{starfield}
\begin{center}
\includegraphics[width=2.5in]{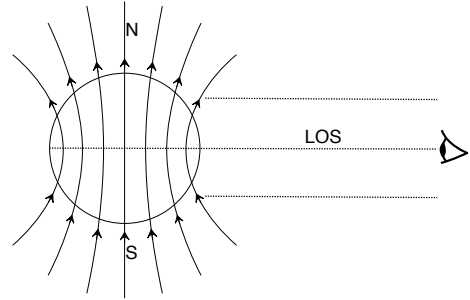}
\caption{Schematic diagram of a star (e.g., Sun) that has a dipolar magnetic
field and is observed (for simplicity) along lines of sight parallel to its 
equatorial plane. To calculate the magnetic flux threading the star by 
observing its {\it surface} field $B_{\rm surf}$, only the mean value of 
$B_{\rm surf}$ in either the northern {\it or} in the southern hemisphere 
should be used. If both values are averaged algebraically (as done by CHT), 
an erroneous flux value of zero will be obtained.}
\end{center}
\vspace{-2ex}
\end{figure}

\subsection{A Self-Consistent Analysis of the CHT Data}

Since both the data and theoretical considerations suggest that $B_{env}$ exhibits 
spatial variations, we reanalyze the CHT data properly accounting for this effect 
and thus generalize the relevance of the data to realistic clouds (instead of idealized 
ones with straight-parallel field lines). High-quality data analyzed in this manner can 
potentially discriminate between alternative fragmentation theories, instead of just 
providing geometrical input to theories.

When intrinsic variation of $B_{\rm los}$ in a cloud's envelope exists, the spread 
in the observed values is the convolution of the measurement error and of 
the intrinsic spread of $B_{\rm los}$. To account for spatial variation of 
$B_{\rm los}$, a likelihood analysis is needed (see Wall \& Jenkins 2003; 
Lyons 1992; Lee 2004). We assume that the ``true'' $B_{\rm los}$ follows a 
Gaussian distribution  with mean $B_0$  and intrinsic spread $\sigma_0$. This 
distribution is then ``sampled'' with $N$ measurements $B_j$, each carrying a 
(Gaussian) error measurement $\sigma_j$. 

\begin{table*}
\begin{center}
\caption{\label{statable} Magnetic Fields and Their Uncertainties (in ${\rm \mu G}$), 
and Upper Limits on the Envelope Field (in ${\rm \mu G}$) and Ratio $R$} 
\begin{tabular}{c|cccc}
\hline \hline
Cloud & $B_{\rm mean} \pm \sigma_{\rm mean} $& $B_{{\rm max}\mathcal{L} }\pm 
\sigma_\mathcal{L}$ & $|B_{\rm env}|(\leq 2\sigma)$ & $|R| (\leq 2\sigma)$\\
\hline
L1448CO& $0\pm 5$ & $-4^{+9}_{-8}$ & $27$ & $2.0$\\
\hline 
B217-2 & $+2 \pm 4$  & $+2 ^{+7}_{-7}$ & $22$  & $2.9$ \\
\hline 
L1544 & $+2\pm 3$   & $+4 ^{+10}_{-8}$  & $29$  & $5.0$\\
\hline 
B1 & $-8 \pm  3$ & $-8^{+5}_{-5}$  & $20$  & $1.1$\\
\hline \hline
\end{tabular}
\end{center}
{\it Col. 1}: The four observed clouds. {\it Col. 2 {\rm \&} 3}: Mean field and 
its uncertainty as given by CHT and by the likelihood data analysis, respectively. 
{\it Col. 4 {\rm \&} 5}: Upper limits on the envelope magnetic field and on the ratio $R$, 
from the likelihood analysis.
\vspace{+6ex}
\end{table*}
At any specific envelope location, there is a probability 
$\exp\left[-(B-B_{0})^2/2\sigma_0^2\right]/\sqrt{2\pi}\sigma_0$ for the magnetic 
field to have a {\em true} value $B$. If the error of measurement
at this same location is $\sigma_j$, then the probability of observing a value $B_j$ 
of the field, {\em given} that its true value is $B$, is  $\exp\left[-(B-B_{j})^2/2\sigma_j^2\right]/\sqrt{2\pi}\sigma_j$. 
However, this is not the only way we could get an observed field value
$B_j$, since there are many different true values of the field that
might yield an observation $B_j$ due to measurement errors. To find the 
{\em total} probability for a single observation of $B_j$, we integrate 
over all possible ``true'' values of the magnetic field at a single location
to get (the likelihood for a single observation $B_j$ with observational 
uncertainty $\sigma_j$):

\begin{equation}\label{like1}
l_{j}= \int_{-\infty}^{\infty} \!\!\!\!\!\!\!dB
\frac{\exp\left[-(B-B_{j})^2/(2\sigma_{j}^2)\right]}
{\sqrt{2\pi}\sigma_{j}}
\frac{ \exp\left[-(B-B_{0})^2/(2\sigma_0^2)\right]}
{\sqrt{2\pi}\sigma_0} \, .
\end{equation}

The likelihood $\mathcal{L}$ for $N$ observations of $B_j$ with individual 
uncertainties $\sigma_j$ to come from an intrinsic probability 
distribution with mean $B_0$ and spread $\sigma_0$ is the product of the 
individual likelihoods, $\mathcal{L} = \prod_{j=1}^Nl_j$ which, after performing 
the integration in equation (\ref{like1}) and some algebraic manipulations, yields 
(see Venters \& Pavlidou 2007)
\begin{equation}\label{likelihood}
\mathcal{L}\left(B_{0}, \sigma_{0}\right) = 
\left(
\prod_{j=1}^{N}\frac{1}{\sqrt{\sigma_0^2+\sigma_{j}^2}}
\right)
\exp\left[-\frac{1}{2} \sum_{j=1}^{N}
\frac{(B_{j} - B_{0})^2 }{\sigma_0^2+\sigma_{j}^2}
\right]\,.
\end{equation}

Any parameters that are not of direct interest (such as the intrinsic spread $\sigma_0$ 
in this case), can be integrated out of the likelihood. In this way, we can derive the 
probability distribution of the parameter of interest ($B_0$) {\em while still allowing 
for all possible values in $\sigma_0$}, rather than arbitrarily demanding that 
$\sigma_0=0$ (as in the CHT analysis). The integrated likelihood is called the 
{\em marginalized likelihood}, $\mathcal{L}_m$; this probability distribution can then 
be used to derive confidence intervals and upper limits where appropriate. The (unnormalized) $\mathcal{L}_m$ for the four clouds is shown in Figures 3a - 3d. $\mathcal{L}_m$ is derived 
by numerically integrating equation (\ref{likelihood}) over $\sigma_0$ for different values 
of $B_0$, and is shown as a solid curve in the four figures; the location of the 
maximum-likelihood estimate for the mean $B_0$ is marked by a heavy vertical line in each 
figure. 

The maximum-likelihood estimates and associated uncertainties of $B_0$ for the four CHT 
clouds are shown in Table \ref{statable}. The uncertainties are systematically greater 
than those quoted by CHT. The $1\sigma$ uncertainties are represented by the widths of the
dark shaded (solid blue) boxes in Figures 3a - 3d. For comparison we show, as cross-hatched 
boxes, the $1\sigma$ spreads of the values of $B_0$ that CHT  quote for the same clouds,  
based on the same data. 

Upper limits for $|B_0|$ can also be calculated using $\mathcal{L}_m$. 
The $2\sigma$ upper limits for the envelope magnetic field  are given 
in Table \ref{statable}. 
(The $2\sigma$ upper limit is that value of $|B_0|$  for which a fractional 
area equivalent  to the Gaussian $2\sigma$ ($95.4\%$) is included under the marginalized 
likelihood curve between $-|B_0|$ and $|B_0|$.  Note that this is {\em not} the 
maximum-likelihood value of $B_0$ plus two times the error. In addition, $\mathcal{L}_m$ 
has much longer tails than a gaussian, and hence $1\sigma$, $2\sigma, ...$ values do not 
scale linearly.) The values of $|B_0|$ and $-|B_0|$ which include between them 
a fractional area of $2\sigma$ of the marginalized likelihood for the four clouds are marked 
with heavy down arrows in Figures 3a - 3d. 

\begin{figure}
\begin{center}
\includegraphics[width=3.5in, clip]{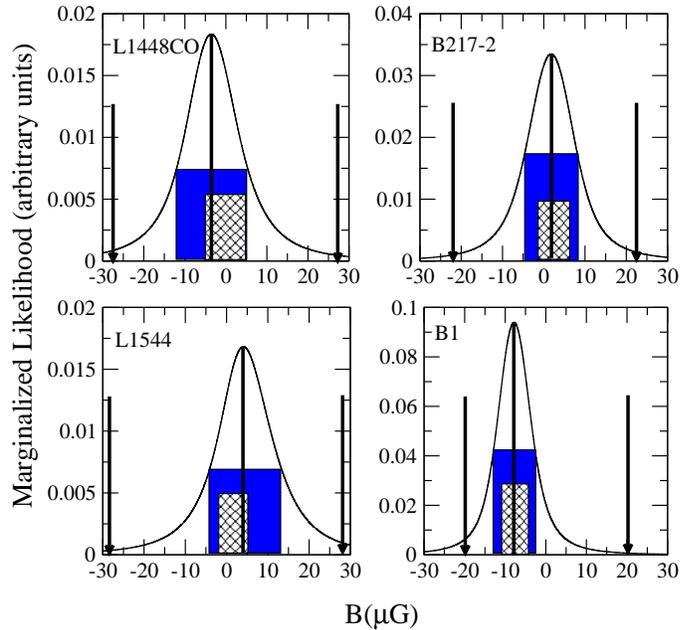}
\caption{Marginalized likelihood (not normalized, solid curve) for the four cloud envelopes 
observed by CHT. In each figure: the location of the maximum-likelihood value of $B_0$ is 
marked with a heavy vertical line; the width of the dark shaded (solid blue) box indicates 
the $1\sigma$ values of $B_0$; the width of the cross-hatched box indicates the $1\sigma$ 
values for $B_0$ according to CHT; heavy down arrows mark the $2\sigma$ upper limit on $|B_0|$. 
\label{marginalized}}
\end{center}
\vspace{+3ex}
\end{figure}

To correctly propagate uncertainties onto the derived quantity $R$, we do a full 
Monte-Carlo calculation to derive the probability distribution for the values of $R$ as 
follows. We repeat the  following experiment $10^6$ times: we draw $I_{\rm core}$, 
$I_{\rm env}$, $\Delta V_{\rm core}$, $\Delta V_{\rm env}$ and $B_{\rm core}$ from 
gaussian distributions with mean and spread equal to the measurement and uncertainty 
quoted in CHT; we draw a mean value of $B_{\rm env}$ from the marginalized likelihood 
of the previous section; we combine all the ``mock observations'' of these numbers to 
produce one value of $R$. We use the $10^6$ values of $R$ produced in this way to 
numerically calculate the probability distribution for $R$. We then calculate the 
$2\sigma$ upper limit on $|R|$ by requiring that the fractional integral of this 
distribution between $-R$ and $+R$ be $95.4\%$. The $2\sigma$ upper limits for $|R|$ 
are given in Table \ref{statable}, last column. These limits are not very strong: 
{\em $|R|$ is constrained to be smaller than a few, and for no cloud is the upper 
limit smaller than $1$} $-$ in sharp contrast to the CHT conclusion, that $R$ is in 
the range 0.02 - 0.42 in the four observed clouds. 

In our analysis, we relaxed only {\it one} of the CHT assumptions (that of lack of 
spatial variation of $B_{\rm env}$, which is not consistent with the data). We have 
retained the implicit assumption of similar orientations of the {\it net} 
${\mathbf B}_{\rm env}$ and ${\mathbf B}_{\rm core}$ ({\it vectors}), because the 
data do not suggest any particular relative orientation of the two vectors. 
A more general analysis that would also relax this assumption would increase the 
uncertainties on $R$ (although not on $B_{\rm env}$) and would further part from 
the CHT conclusions.

\section{Summary and Conclusion}\label{future}

We have presented a self-consistent analysis of recent OH Zeeman observations by Crutcher 
\etal (2009), and have shown how such data can be combined in a statistically robust 
manner to obtain constraints on the mass-to-flux ratio contrast ($R$) between molecular 
cloud cores and envelopes. Our analysis extends the constraining power of such measurements 
to realistic clouds, beyond the overly restrictive assumption of straight-parallel field 
lines in cloud envelopes, adopted by CHT. We have shown that the CHT data are not of 
good enough quality to constrain the ratio $R$ and thereby test molecular-cloud 
fragmentation theories: (i) more integration time is needed to reduce measurement errors 
(now most measurements of $B_{\rm env}$ yield only upper limits); and (ii) more 
measurements (instead of only four) in each envelope are needed to better constrain 
the intrinsic spatial variation of $B_{\rm env}$. This kind of observations 
coupled with the method of data analysis we presented in this {\it Letter} has great 
promise and can lead to significant progress in the field of ISM physics in general and in 
understanding the role of magnetic fields in molecular-cloud dynamics in particular.

\section*{Acknowledgements}
We thank Robert Dickman, Paul Goldsmith, Mark Heyer, Dan Marrone, and Vasiliki Pavlidou for
valuable discussions. 
TM acknowledges partial support by NSF under grant AST-07-09206, and KT by JPL/Caltech, 
under a contract with the National Aeronautics and Space Administration. \copyright 2009. 
All rights reserved.

\end{document}